\documentclass[
reprint,
amsmath, amssymb,
aps, floatfix
pra,
]{revtex4-2}

\usepackage{graphicx}
\usepackage{dcolumn}
\usepackage{bm}
\usepackage[colorlinks]{hyperref}
\usepackage{braket}
\usepackage{physics}
\usepackage{amsmath}
\usepackage{array}
\usepackage{float}

\usepackage[normalem]{ulem}

\begin{document}

\title{Periodic-Pulse Control of Collective Emission and Photon Statistics in Interacting Quantum Emitters}


\author{ Kavalambramalil George Paulson, Fu Li, and Hanna Terletska}
\affiliation{Department of Physics and Astronomy, Middle Tennessee State University, Murfreesboro, TN 37132, USA}
\affiliation{Quantum Research, Interdisciplinary Science, and Education (QRISE) Center, Middle Tennessee State University, Murfreesboro, TN 37132, USA}

\email{ paulson.kavalambramalilgeorge@mtsu.edu}




\date{\today}

\begin{abstract}

We investigate the collective emission spectra and photon correlations of two identical dipole-dipole-coupled emitters under continuous-wave (CW) and periodic pulsed driving. Our results show that a periodic train of instantaneous $\pi_x$ pulses can generate emission spectra whose dominant features closely resemble those obtained under CW excitation.  We find that the structure of the collective emission spectra reflects the interplay among the resonance conditions, dipole-dipole coupling, and the inter-pulse delay $\tau_D$.  We further demonstrate that finite-duration $\pi_x$ pulses with sufficiently high Rabi frequencies can closely reproduce the ideal instantaneous-pulse limit. In contrast to the emission spectra, the photon correlations $g^{(2)}(t,\tau)$ are governed primarily by $\tau_D$ and, unlike under CW driving, become increasingly insensitive to the resonance conditions in the short-pulse limit. These results establish periodic pulsed driving as a versatile means of tailoring both the emission spectra and the photon statistics of coupled two-emitter systems.

\end{abstract}

\maketitle


\section{\label{sec:level1} Introduction}
Interactions between light and matter underpin quantum optical science and technology, giving rise to processes such as absorption, emission, and scattering, as well as coherent, nonlinear, and collective effects~\cite{scully1997quantum}. These mechanisms enable the preparation, control, and measurement of quantum states, forming the basis of technologies in sensing, communication, and information processing~\cite{kimble2008quantum,o2009photonic}.  One of the most well-investigated models in quantum optics is the two-level quantum emitter driven by a coherent electromagnetic field~\cite{mollow1969power,kimble1977photon}, which provides a basic framework for single-photon generation and related quantum technologies, including quantum communication, computation, sensing, and quantum energy-storage models~\cite{aharonovich2016solid,senellart2017high,couteau2023applications,degen2017quantum,campaioli2024colloquium,george2025work}.

The dynamical response of such driven quantum systems is directly reflected in their spectral properties. In particular, the emission and absorption spectra 
encode information about coherence, population dynamics, and system-field coupling. The spectral and statistical properties of the emitted light depend sensitively on the nature of the driving field and on the relevant interaction strengths. For a single driven emitter, the celebrated Mollow triplet in the steady-state fluorescence spectrum~\cite{mollow1969power}  and photon antibunching~\cite{kimble1977photon} in the second-order intensity correlation function are hallmark features that have been extensively studied under continuous-wave excitation. 
When several emitters are considered, collective behavior arises from radiative dipole-dipole interactions mediated by the electromagnetic field. In closely spaced systems, these interactions induce both coherent energy shifts and collective decay channels, giving rise to superradiant and subradiant modes that strongly influence the emission spectrum and photon statistics. Collective radiative effects, first predicted in Dicke’s theory~\cite{dicke1954coherence} of cooperative spontaneous emission, are now investigated in a variety of systems~\cite{scheibner2007superradiance, mlynek2014observation, angerer2018superradiant} where sub‑wavelength emitter spacing enables modified emission dynamics and collective decay channels.  In the two-emitter case, these interactions split the collective single-excitation states and generate sidebands to the cooperative resonance-fluorescence
spectrum~\cite{freedhoff1979collective,agarwal1980analytical, Griffin1982, Compagno1982}, while also reshaping the photon statistics of the emitted light~\cite{FicekTanasKielich1984,beige1998transition,darsheshdar2021photon,vivas2021two,juan2024tailoring}.

Although Dicke emphasized the intrinsically transient nature of cooperative decay~\cite{dicke1954coherence}, much of the subsequent theoretical literature has focused on steady-state regimes~\cite{freedhoff1979collective,agarwal1980analytical, Griffin1982, Compagno1982, FicekTanasKielich1984,beige1998transition,darsheshdar2021photon,vivas2021two,juan2024tailoring}, in which continuous driving and stationary populations render the analysis more tractable.
The dynamics of coupled emitters, however, are inherently time-dependent, governed by the buildup and decay of multi-excitation coherence on the timescale of the natural lifetime. Transient spectra and photon correlations, both under CW driving before the steady state is reached, and under pulsed driving before the long-time limit, carry signatures qualitatively distinct from, and inaccessible to, stationary descriptions~\cite{eberly1980time,ho1988time,moelbjerg2012resonance,vinas2020time,koong2022coherence,fotso2023tuning}.
These transient regimes are now experimentally accessible: recent experiments in solid-state, atomic, and superconducting platforms have demonstrated controlled single- and multi-emitter quantum-optical dynamics, including resonance fluorescence and collective radiative effects such as superradiance and subradiance~\cite{mlynek2014observation,guerin2016subradiance,kim2018super,tiranov2023collective}.
In these systems, time-resolved measurements have directly probed the buildup and decay of cooperative emission~\cite{guerin2016subradiance,kim2018super,okaba2019superradiance,tiranov2023collective}. These advances motivate theoretical studies of transient emission spectra and two-time photon correlations under periodic pulsed driving, where cooperative dynamics are encoded in the temporal and spectral structure of the emitted light. The regime combining dipole-dipole coupling with pulse sequences, however, remains considerably less explored than its CW steady-state counterpart~\cite{darsheshdar2021photon,vivas2021two,bardetski2024dipole}.

Pulsed driving offers a complementary handle on these dynamics. Ultrafast optical control of quantum-dot spins and excitons~\cite{press2008complete,wigger2018rabi}, single-photon source characterization under pulsed excitation~\cite{Fischer2017, Gustin2018}, and periodic-pulse control of the emission spectrum of a single emitter via the inter-pulse delay~\cite{fotso2019controlling, Wang2026} have all been demonstrated. Here, we extend this pulse-delay spectral control to a dipole-dipole-coupled pair, where pulse-induced sidebands superimpose on the collective transition frequencies, a regime not addressed by single-emitter pulse studies~\cite{fotso2019controlling, Wang2026} nor by steady-state collective treatments such as the $N$-emitter CW analysis of~\cite{bardetski2024dipole}

Building on these developments, we investigate two identical emitters coupled by dipole-dipole interaction and driven either by a CW field or by a periodic train of instantaneous $\pi_x$ pulses. We consider both single- and two-photon resonance and span the weak- and strong-coupling regimes. Under CW driving, the collective emission evolves from a Mollow-like triplet at weak dipole-dipole coupling to a resolved multi-peak structure at strong coupling and strong driving, which we interpret using the dressed-state basis of the coupled atom-field system. One of our main results is that,  with a appropriately chosen inter-pulse delays, periodic pulsed driving  produces dominant spectral features that qualitatively resemble those obtained under CW excitation. The inter-pulse delay $(\tau_D)$ sets the spacing of pulse-induced sidebands around the collective transition frequencies and thereby provides a controllable parameter for shaping the transient spectrum.

We further compute the transient two-time correlation $g^{(2)}(t,\tau)$ for both driving schemes. Under CW excitation, the photon statistics are strongly shaped by the resonance condition. Under pulsed driving, this dependence is reduced and vanishes entirely in the instantaneous-pulse limit, where the statistics are instead set by the inter-pulse delay, a regime of practical relevance to on-demand single-photon source engineering~\cite{Fischer2017, Gustin2018}. Finally, we validate the instantaneous-pulse approximation through an average-gate-fidelity analysis, showing that finite-duration pulses of
sufficiently high Rabi frequency reproduce the idealized impulsive limit. Taken together, these results identify the inter-pulse delay as the common control parameter for both the collective emission spectrum and the transient photon statistics of a coupled-emitter pair as the pulses approach the instantaneous limit.

The paper is organized as follows. Section II introduces the theoretical model and master-equation framework. Section III presents the CW fluorescence spectrum. Section IV analyzes transient emission under pulsed excitation and its correspondence to the CW case, including a finite-duration pulse-fidelity analysis. Section V examines transient photon–photon correlations for both driving schemes. Section VI discusses our results and summarizes the findings.

\section{Model}
We consider a system comprising two two-level atoms interacting via dipole-dipole interactions and driven by an external coherent laser field. The Hamiltonian describes the system
\begin{align}
\nonumber
\hat{H}_s = \hbar \sum_{i=1}^{2} \omega_i  s_i^z + \hbar \sum_{i \neq j}^{2} \Omega_{ij} s_i^+ s_j^- \\
 -\frac{1}{2} \hbar \sum_{i=1}^{2} \left[ \Omega(\vec{r}_i) s_i^+ e^{-i (\omega_L t + \phi_L)} + \text{H.c.} \right].
\label{Hamiltonian_Orig}
\end{align}
where,
${s_i}^+ = \vert e_i\rangle \langle g_i\vert$, ${s_i}^- = \vert g_i\rangle \langle e_i\vert$, $s_i^z = \frac{1}{2} \left( |e_i\rangle \langle e_i| - |g_i\rangle \langle g_i| \right)$, are
respectively, the raising, lowering, and inversion operators of the $i^{\text{th}}$ atom; $\omega_i$ is the transition frequency of atom $i$;
$\Omega_{ij}$ is the dipole-dipole coupling strength between the atoms. The laser field has frequency $\omega_L$ and phase $\phi_L$. The driving strength is set by the Rabi frequency $\Omega(\vec{r_i})$, which at the position of the atom $i$ is
\begin{equation}
\Omega({\vec{r_i}})\equiv\Omega_i=\frac{\vec{\mu_i}.\vec{E_L}e^{i\vec{k_L}.\vec{r_i}}}{\hbar},
\end{equation}
where $\vec{\mu_i}$ is the atomic transition dipole  moment, $\vec{E_L}=E_0 \hat{\epsilon}$  is the complex field amplitude vector, and $\vec{k_L}$ is the wave vector of the driving field, respectively~\cite{ficek2002entangled}.

\begin{figure}[bt]
\centering
\includegraphics[height=50mm,width=\linewidth]{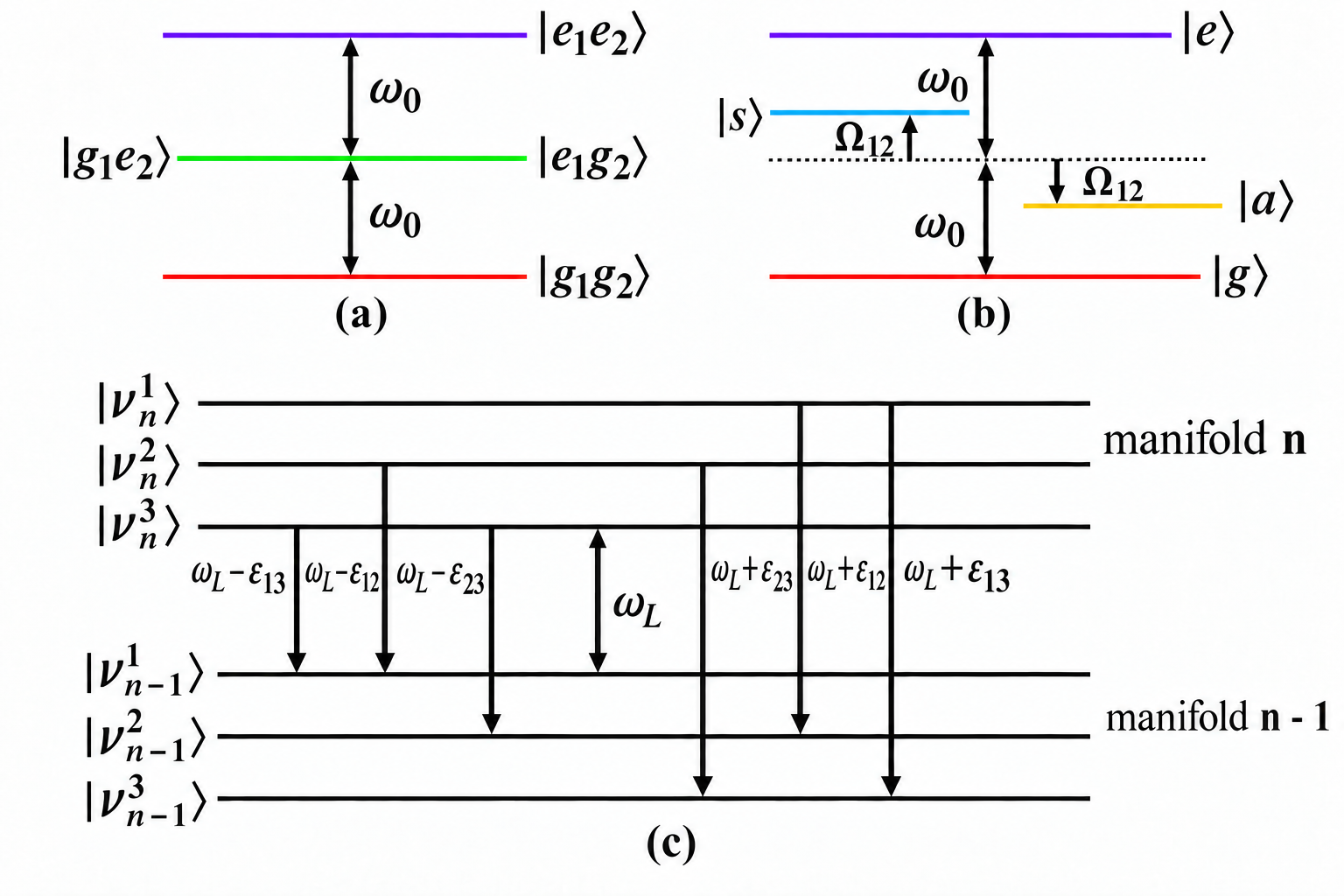}
\caption{Energy level diagrams of an identical two-emitter system in the bare, collective, and dressed bases are shown in Figs(a), (b), and (c), respectively. }
\label{Energy_Level_Diagram}
\end{figure}

We consider two identical atoms with $\omega_1 = \omega_2 = \omega_0$. In the absence of an external driving field and dipole-dipole interaction, the singly-excited states $\lvert g_1 e_2 \rangle$ and $\lvert e_1 g_2 \rangle$ are degenerate (Fig.~\ref{Energy_Level_Diagram}(a)). Together with $\lvert g_1 g_2 \rangle$ and $\lvert e_1 e_2 \rangle$, they form the bare-state basis, with eigenvalues $E_{gg} = -\hbar \omega_0$, $E_{ge} = E_{eg} = 0$, and $E_{ee} = \hbar \omega_0$.

The dipole-dipole coupling $\Omega_{12} = \Omega_{21} \neq 0$, lifts the degeneracy of the singly-excited states. The eigenvalues become $E_G = -\hbar \omega_0$, $E_A = -\hbar \Omega_{12}$, $E_S = \hbar \Omega_{12}$, and $E_E = \hbar \omega_0$ (Fig.~\ref{Energy_Level_Diagram}(b)), with the corresponding collective eigenstates 
\begin{align}
\nonumber
	\vert E\rangle &= \vert e_1 e_2\rangle, \\ \nonumber
	\vert S\rangle &= \frac{1}{\sqrt{2}}\bigl(\vert e_1 g_2\rangle + \vert g_1 e_2\rangle\bigr), \\ \nonumber
	\vert A\rangle &= \frac{1}{\sqrt{2}}\bigl(\vert e_1 g_2\rangle - \vert g_1 e_2\rangle\bigr), \\
	\vert G\rangle &= \vert g_1 g_2\rangle.
\label{collective_bais}
\end{align}
The symmetric state $\lvert S \rangle $  and antisymmetric state $\lvert A \rangle$ are separated in energy by $2\hbar\Omega_{12}$. This collective basis is the natural one for the interacting system, since it diagonalizes the dipole-dipole coupling and makes super- and subradiant decay channels explicit; we therefore work in this basis. The unitary transformation that maps the product-state basis of the two atoms onto the collective (symmetric-antisymmetric) basis is given by
\begin{equation}
U =
\begin{bmatrix}
1 & 0 & 0 & 0 \\
0 & -\frac{1}{\sqrt{2}} & \frac{1}{\sqrt{2}} & 0 \\
0 & \frac{1}{\sqrt{2}} & \frac{1}{\sqrt{2}} & 0 \\
0 & 0 & 0 & 1
\end{bmatrix}.
\end{equation}


Since the driving field propagates perpendicular to the interatomic axis, $\hat{k}_L \perp \hat{r}_{12}$, the positional phase factor satisfies 
$e^{i\vec{k}_L\cdot\vec{r}_{12}} = 1$. Consequently, both emitters experience the same Rabi frequency, $\Omega_0$, independent of their separation. From Eq.~\ref{Hamiltonian_Orig},
for two identical atoms, $\lvert\vec{\mu}_1\rvert=\lvert\vec{\mu}_2\rvert$, driven by a coherent laser field with equal driving strengths, $\Omega(r_1)=\Omega(r_2)=\Omega_0$, and setting the laser phase $\phi_L=0$ without loss of generality, the Hamiltonian becomes

\begin{align}
\hat{H_s}&=\hbar\omega_0( s_1^{z}+ s_2^z) +\hbar\sum_{i\neq j=1}^2\Omega_{ij}s_{i}^{+}s_{j}^{-}\nonumber\\
&\quad-\frac{1}{2}\hbar\Omega_0[(s_1^{+}+ s_2^{+})e^{-i\omega_L t}+(s_1^{-}+s_2^{-})e^{i\omega_L t}].
\end{align}

Transforming to the frame rotating at the laser frequency $\omega_L$ via $U_{R} = e^{i \omega_L (s_1^{z} + s_2^{z}) t}$, the explicit time dependence of the Hamiltonian is removes and $\hat{H_s}$ takes the time-independent form
\begin{align}
\hat{H_s^{'}} &= \hbar \Delta( s_1^z + s_2^z)
+ \hbar\sum_{i\neq j=1}^2\Omega_{ij}s_{i}^{+}s_{j}^{-} \nonumber \\
&\quad -\frac{1}{2} \hbar\Omega_0 (s_1^+ + s_2^+ + s_1^- + s_2^-),
\end{align}
where $\Delta=\omega_0-\omega_L$ is the atom-laser detuning.

The master equation for the two-atom system, driven by an external coherent laser field and coupled to a common vacuum electromagnetic field, under the Born-Markov approximation, is given by~\cite{ficek2002entangled,almutairi2011generating}
\begin{equation}
\frac{\partial \hat{\rho}}{\partial t} = -\frac{i}{\hbar} [\hat{H}_s^{'}, \hat{\rho}] - \frac{1}{2} \sum_{i,j=1}^{2} \Gamma_{ij} \left( \hat{\rho} s_i^+ s_j^- + s_i^+ s_j^- \hat{\rho} - 2 s_j^- \hat{\rho} s_i^+ \right).
\end{equation}
Substituting the values of $\hat{H}_s^{'}$ and setting $\hbar=1$, the master equation takes the form
\begin{align}
\dot{\rho} = & -i\Delta \left[ s_1^z + s_2^z, \rho \right] 
- i \sum_{i \neq j = 1}^{2} \Omega_{ij} [s_i^+ s_j^-, \rho] \nonumber \\
& + \frac{i \Omega_0}{2} \sum_{i=1}^{2} [s_i^+ + s_i^-, \rho] \nonumber \\
& - \frac{1}{2} \sum_{i,j=1}^{2} \Gamma_{ij} \left( \rho s_i^+ s_j^- + s_i^+ s_j^- \rho - 2 s_j^- \rho s_i^+ \right).
\label{ME_Eqn}
\end{align}


The dissipative part of the master equation~(\ref{ME_Eqn}), governed by the rates $\Gamma_{ij}$, describes the evolution of $\hat{\rho}$ due to spontaneous emission. The diagonal element $\Gamma_{ii}\equiv\Gamma$ is the single-atom spontaneous emission rate. For identical atoms, it is independent of $i$, so both atoms decay at the same rate,
\begin{equation}
\Gamma = \frac{\omega_0^3\mu^2}
{3\pi\epsilon_0\hbar c^3}.
\end{equation}
The off-diagonal element $\Gamma_{ij}$ ($i\neq j$) is the collective spontaneous emission rate arising from the coupling of both atoms to the common vacuum electromagnetic field,
\begin{equation}
\Gamma_{ij} =\sqrt{\Gamma_i \Gamma_j}\, F(k_0 r_{ij})\equiv \Gamma\,F(k_0 r_{ij}), \qquad (i \neq j),
\end{equation}
for identical atoms $\Gamma_i=\Gamma_j\equiv\Gamma$, and $F(k_0 r_{ij})$ is a dimensionless geometric factor encoding the orientational and distance dependence of the interatomic radiative coupling:

\begin{widetext}
\begin{align}
\small
F(k_0 r_{ij}) = \frac{3}{2} \bigg\{  
    & \left[(\hat{\mu}_i\cdot\hat{\mu}_j) - (\hat{\mu}_i \cdot \hat{r}_{ij}) (\hat{\mu}_j \cdot \hat{r}_{ij}) \right]  
    \frac{\sin(k_0 r_{ij})}{k_0 r_{ij}} \nonumber \\  
    & + \left[(\hat{\mu}_i\cdot\hat{\mu}_j) - 3 (\hat{\mu}_i \cdot \hat{r}_{ij}) (\hat{\mu}_j \cdot \hat{r}_{ij}) \right]  
    \left[ \frac{\cos(k_0 r_{ij})}{(k_0 r_{ij})^2}  
    - \frac{\sin(k_0 r_{ij})}{(k_0 r_{ij})^3} \right]  
    \bigg\}.
\label{F_value}
\end{align}

Dipole-dipole coupling between atoms is given as
\begin{align}
\small
    \Omega_{ij}= \frac{3}{4} \sqrt{\Gamma_i \Gamma_j} \bigg\{ 
    & -\left[(\hat{\mu}_i\cdot\hat{\mu}_j) - (\hat{\mu}_i \cdot \hat{r}_{ij})(\hat{\mu}_j \cdot \hat{r}_{ij}) \right] \frac{\cos(k_0 r_{ij})}{k_0 r_{ij}} \nonumber \\
    & + \left[(\hat{\mu}_i\cdot\hat{\mu}_j) - 3 (\hat{\mu}_i \cdot \hat{r}_{ij})(\hat{\mu}_j \cdot \hat{r}_{ij}) \right] 
    \left[ \frac{\sin(k_0 r_{ij})}{(k_0 r_{ij})^2} + \frac{\cos(k_0 r_{ij})}{(k_0 r_{ij})^3} \right] 
    \bigg\}.
    \label{Omega_12vv}
\end{align}
\end{widetext}

With the geometry established above, $\hat{\mu}_1=\hat{\mu}_2\equiv\hat{\mu}$ and $\hat{\mu}\perp\hat{r}_{12}$, the dot products entering $F$ and $\Omega_{12}$ simplify to
$\hat{\mu}_i\cdot\hat{\mu}_j=1$ and $\hat{\mu}_i\cdot\hat{r}_{12}=0$, yielding the expressions given in Eqs.~(\ref{F_value}) and~(\ref{Omega_12vv}), where $k_0=\omega_0/c=2\pi/\lambda$ is the resonant wave vector and $\lambda$ is the resonant wavelength. For large separations, $r_{12}\gg\lambda$, the envelopes of both $\Gamma_{12}$ and $\Omega_{12}$ decay as $(k_0r_{12})^{-1}$, and the corresponding collective effects become negligibly small. In contrast, collective effects become significant for $r_{12}\lesssim\lambda/2$. Together, the collective coupling modifies the spontaneous emission and induces a coherent interaction between the emitters. In the following section, we examine how the coupling strength and the
driving field shape the collective fluorescence of the two-emitter system.

\section{Fluorescence  Spectrum Under Continuous Wave Driving}

We next characterize the emission of the driven two-emitter system through its fluorescence spectrum, which provides direct access to the spectral structure of the emitted radiation.
The emission spectrum associated with the transition operators
\(a_{jk}^{+} = |j\rangle\langle k|\) and \(a_{jk}^{-} = |k\rangle\langle j|\), for \(j \neq k\), is given by~\cite{mollow1972absorption,mollow1972stimulated}
\begin{equation}
\mathcal{S}_1(\omega) = \int_0^T dt \int_0^{T - t} d\theta \,
\langle a_{jk}^+(t + \theta)\, a_{jk}^-(t) \rangle \, e^{-i\omega\theta}.
\label{E_spectrum}
\end{equation}
The quantity \(S(\omega) = 2A^2\,\mathrm{Re}\{\mathcal{S}_1(\omega)\}\) 
gives the system's direct emission spectrum, where $A$ is a real constant determined by the detection geometry and does not affect the spectral shape; we set $A=1$. 

To evaluate the emission spectrum, we apply the quantum regression theorem and express the two-time correlation function in terms of an auxiliary density matrix,
\begin{equation}
\langle a_{jk}^{+}(t+\theta)\, a_{jk}^{-}(t) \rangle = \mathrm{Tr}\!\left[ a_{jk}^{+}\, \rho'(t,t+\theta) \right],
\end{equation}
where $a_{jk}^{+} = |j\rangle\langle k|$ and $a_{jk}^{-} = |k\rangle\langle j|$ are the transition operators between the collective states. The auxiliary density matrix $\rho'(t,t+\theta)$ is obtained by evolving the initial condition
\begin{equation}
\rho'(t,t) = a_{jk}^{-}\, \rho(t)
\end{equation}
forward in time from $t$ to $t+\theta$ according to the master equation~(Eq.~\ref{ME_Eqn}), including all pulses applied within this interval. The spectrum contains contributions from transitions originating in both the singly excited states ($S, A \to G$) and the doubly excited state ($E \to S, A$), which are obtained through the corresponding choices of $a_{jk}^{\pm}$.


For the case of pulsed driving, the pulses are assumed to be strong and short, such that $\Omega_0 \gg \Gamma, \Delta$, and can thus be treated as effectively instantaneous. At the pulse times $t=n\tau$, the density matrix undergoes a unitary transformation,
\begin{equation}
\rho(n\tau^+) = U\,\rho(n\tau^-)\,U^\dagger,
\end{equation}
where $U$ represents the unitary operator corresponding to the pulse applied to both atoms. Here,
\begin{equation}
\rho(n\tau^\pm)
\equiv
\lim_{\epsilon\rightarrow0^+}\rho(n\tau\pm\epsilon),
\end{equation}
with $\rho(n\tau^-)$ and $\rho(n\tau^+)$ denoting the density matrices immediately before and after the pulse, respectively, $n$ being an integer, and $\tau$ denoting the inter-pulse delay.




\subsection{Emission Spectra Across Different Coupling Regimes}


We study the transient collective fluorescence of two identical coupled emitters under continuous symmetric excitation.  The emission spectra are controlled by the interplay of the dipole-dipole interaction $\Omega_{12}$ and the driving strength $\Omega_0$, which we explore across weak-coupling ($|\Omega_{12}| < \Gamma$) and strong-coupling ($|\Omega_{12}| \gg \Gamma$) regimes under moderate ($\Omega_0 \sim \Gamma$) and strong ($\Omega_0 \gg \Gamma$) driving.
In the collective basis, decay from the doubly excited state $\vert E\rangle$ can occur through the symmetric superradiant channel $\vert E\rangle \to \vert S\rangle \to \vert G\rangle$, with decay rate $\Gamma+\Gamma_{12}$, or through the antisymmetric subradiant channel $\vert E\rangle \to \vert A\rangle \to \vert G\rangle$, with decay rate $\Gamma-\Gamma_{12}$ { (Fig.~\ref{Energy_Level_Diagram} b)}. Since the identical emitters are driven symmetrically, the laser couples only to the bright ladder $\vert G\rangle \leftrightarrow \vert S\rangle \leftrightarrow \vert E\rangle$, selected by $S^-=\sigma_1^-+\sigma_2^-$, whereas $\vert A\rangle$ remains dark to the drive. We therefore restrict our analysis to the transient fluorescence of the superradiant pathway and neglect the antisymmetric contribution.

\begin{figure}[h!]
\centering
\includegraphics[height=70mm,width=\linewidth]{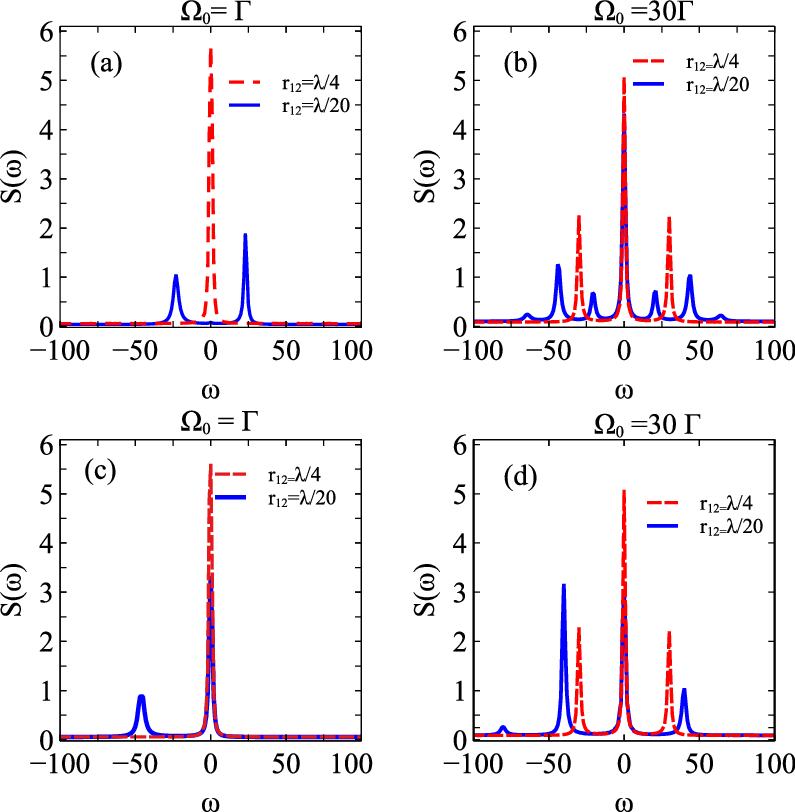}
\caption{Collective emission spectrum from a coupled two identical ($\Delta_1=\Delta_2$) emitter system under continuous driving fields for $\hat{\mu}\perp\hat{r}_{12}$. In  (a) and (b), the laser is tuned to two-photon resonance ($2\omega_0=2\omega_L$) with Rabi frequencies $\Gamma$ and $30 \Gamma$, respectively. The laser is tuned to the transition frequency of the superradiant state ($\Delta=-\Omega_{12}$) in  (c) and (d), with the Rabi frequency set to $\Gamma$ and $30\Gamma$, respectively. The cases of weak and strong dipole–dipole coupling are considered, corresponding to $r_{12} = \lambda/4$ and $r_{12} = \lambda/20$, where $r_{12}$ and $\lambda$ denote the interatomic distance and the resonant wavelength, respectively. The total time of evolution is considered to be $\Gamma t=4$.}
\label{fig2_CW}
\end{figure}

In Fig.~\ref{fig2_CW}, we consider the configuration $\hat{\bm{\mu}}\perp\hat{\bm{r}}_{12}$ under continuous-wave (CW) driving for different coupling and driving regimes. The atoms are 
initially prepared in the fully excited state $|e_1 e_2\rangle$. Two inter-emitter separations are considered: $r_{12}=\lambda/4$, corresponding to weak dipole-dipole coupling 
($|\Omega_{12}|\lesssim\Gamma$), and $r_{12}=\lambda/20$, corresponding to strong coupling ($|\Omega_{12}|\gg\Gamma$). In Figs.~\ref{fig2_CW}(a) and~\ref{fig2_CW}(b), the laser is 
tuned to two-photon resonance, $2\omega_L = E_E - E_G$ (i.e., $\Delta = 0$); in Figs.~\ref{fig2_CW}(c) and~\ref{fig2_CW}(d), it is tuned to the single-photon resonance of 
the superradiant transition, $\Delta = -\Omega_{12}$. Within each resonance condition, we compare moderate driving ($\Omega_0 \sim \Gamma$) with strong driving ($\Omega_0 \gg \Gamma$), 
yielding four distinct regimes explored in panels (a)-(d).

In the weak-coupling, moderate-driving regime, $\vert \Omega_{12}\vert<\Gamma\sim\Omega_{0}$, the fluorescence spectrum is dominated by a central peak at $\omega=0$ (red curve in Fig.~\ref{fig2_CW}(a)), arising mainly from coherent fluorescence, as the weak dipole-dipole interaction induces only a small splitting between the symmetric ($\vert S\rangle$) and antisymmetric ($\vert A\rangle$) states, so that no resolved sidebands appear. In the weak-coupling, strong-driving regime ($\Omega_{0}=30\Gamma$), $\vert\Omega_{12}\vert<\Gamma<<\Omega_0$ (red curve in Fig.~\ref{fig2_CW}(b)), the fluorescence spectrum exhibits a Mollow-like triplet, with a central peak at $\omega=0$ corresponding to coherent fluorescence and sidebands at $\omega\approx\pm\Omega_{0}$, as the strong driving field sets the dominant spectral splitting while the dipole-dipole interaction remains weak. In contrast, for strong coupling ($r_{12}=\lambda/20$) and moderate driving ($\vert\Omega_{12}\vert>>\Omega_0\sim\Gamma$), the dipole-dipole interaction strongly splits the collective one-excitation states, so that sidebands dominate the fluorescence spectrum at $\omega\approx\pm\Omega_{12}$ (blue curve in Fig.~\ref{fig2_CW}(a)), while central contribution remains comparatively weak. The unequal heights of the two peaks reflect the population imbalance between the collective states.
  
For weak coupling, the single-photon-resonance spectra (Figs.~\ref{fig2_CW}(c), (d), red curves) remain qualitatively similar to those at two-photon resonance (Figs.~\ref{fig2_CW}(a), (b), red curves) at both driving strengths; the small dipole-dipole splitting renders the spectra relatively insensitive to the choice of resonance condition. In strong coupling, the single-photon resonance noticeably modifies the spectrum. At moderate driving, sidebands appear around the shifted resonance, consistent with emission through
the superradiant channel (Fig.~\ref{fig2_CW}(c), blue curve).  
\subsection{Dressed-State Interpretation of the emission spectrum}
Although the collective-state basis suffices to describe the dipole-dipole-induced splitting between the symmetric and antisymmetric states, it cannot account for the multi-peak fluorescence spectrum that emerges in the strong-coupling and strong-driving regime. A physically transparent description is provided by the dressed-state picture of the driven two-emitter system, in which spectral features arise from transitions between collective atom-field dressed states. 

To formalize this description, we construct the dressed states of the coupled atom-field system in a joint atom-photon basis.  In this fully quantized atom-field   framework, the total excitation operator
\begin{equation}
	N_E = N_k + \sum_{i=1,2} \sigma_i^{+}\sigma_i^{-},
\end{equation}
where $N_k = a^\dagger a$ is the photon number operator, commutes with the full atom-field Hamiltonian under the rotating-wave approximation, $[H, N_E]=0$. Since $N_E$ is conserved, its eigenvalue $n$ labels invariant excitation manifolds. The four-dimensional subspace corresponding to eigenvalue $n$ is spanned by the collective atom-photon product states
\begin{align}
\nonumber
	\chi^{n}_{1} &= \lvert E,\, n-2 \rangle, \\ \nonumber
	\chi^{n}_{2} &= \lvert S,\, n-1 \rangle, \\ \nonumber
	\chi^{n}_{3} &= \lvert A,\, n-1 \rangle, \\
	\chi^{n}_{4} &= \lvert G,\, n \rangle,
\end{align}
with the collective atomic states $\vert E\rangle$, $\vert S\rangle$, $\vert A\rangle$ and $\vert G\rangle$  defined in Eq.~\ref{collective_bais}.
Diagonalization of the atom-field Hamiltonian within this four-dimensional $n$-excitation manifold yields the dressed eigenstates. For identical emitters under equal driving ($\Omega_1 = \Omega_2 \equiv \Omega_0$), the Hamiltonian restricted to the $n$-excitation manifold takes the form
\begin{align}
	\vert \nu_n^1\rangle &=
	\frac{1}{\sqrt{2+\left(\frac{k_+}{g}\right)^2}}
	\left(
	\vert E,n{-}2\rangle
	-\frac{k_+}{g}\,\vert S,n{-}1\rangle
	+\vert G,n\rangle
	\right), \\
	\vert \nu_n^0\rangle &= \vert A,n{-}1\rangle, \\
	\vert \nu_n^2\rangle &=
	\frac{1}{\sqrt{2}}\bigl({-}\vert E,n{-}2\rangle
	+\vert G,n\rangle\bigr), \\
	\vert \nu_n^3\rangle &=
	\frac{1}{\sqrt{2+\left(\frac{k_-}{g}\right)^2}}
	\left(
	\vert E,n{-}2\rangle
	-\frac{k_-}{g}\,\vert S,n{-}1\rangle
	+\vert G,n\rangle
	\right),
\end{align}
where
\begin{equation}
k_{\pm}=\frac{\Omega_{12}\pm\sqrt{\Omega_{12}^2+8g^2}}{2},	\qquad
g=\frac{\Omega_0}{\sqrt{2}}.
\end{equation}

All dressed states in the \(n\)-excitation manifold are atom-field entangled, except \(\nu_n^{0}\). For identical emitters under symmetric driving, $\vert A\rangle$ is dark to the drive; we therefore focus on the transient fluorescence of the bright channel.

The dressed states \(\{\nu_n^{i}\}\), \(i=1,2,3\), form the \(n\)-excitation manifold, with intramanifold level splittings \(\epsilon_{ij} = E_n^{i} - E_n^{j}\). The dressed-state spectrum forms a ladder of such excitation manifolds, each separated by \(\omega_L\). Single-photon emission arises from transitions between adjacent manifolds,
\begin{equation}
	\vert \nu_n^i\rangle \rightarrow \vert \nu_{n-1}^j\rangle,
\end{equation}
reflecting the loss of one excitation between manifolds.

In the strong-coupling, strong-driving regime $(\Omega_0 > |\Omega_{12}| \gg \Gamma)$ (Fig.~\ref{fig2_CW}(b)), the emission spectrum (blue curve) is governed by transitions between adjacent manifolds. The diagonal transitions $(\vert \nu_n^i\rangle \to \vert \nu_{n-1}^i\rangle)$ $(i=1,2,3)$ yield the central spectral component at $\omega=\omega_L$, while off-diagonal transitions $\vert \nu_n^i\rangle \to \vert \nu_{n-1}^j\rangle$$(i \neq j)$ (Fig.~\ref{Energy_Level_Diagram}(c)) generate six sidebands at frequencies $\omega-\omega_L=\pm \epsilon_{ij}$. Thus, at the two-photon resonance, these adjacent-manifold transitions produce a seven-component fluorescence spectrum consisting of one central peak and six sidebands. Away from the two-photon resonance, the dressed-state energies and transition strengths are modified, and the spectral components need not retain this symmetric seven-peak structure.

For comparison, in the single-emitter limit, the four-state manifold collapses to the dressed doublet
\begin{equation}
\vert \pm, n\rangle	= 	\frac{1}{\sqrt{2}} 	\bigl( 	\vert e, n{-}1\rangle \pm\vert g,n\rangle 	\bigr),
\end{equation}
whose inter-manifold transitions produce the Mollow triplet of a strongly driven quantum emitter: a central peak at the laser frequency ($\omega_L$) from the diagonal transitions $|\pm,n\rangle\to|\pm,n{-}1\rangle$, and two symmetric sidebands at $\omega_L\pm\Omega_0$ from the off-diagonal transitions $|\pm,n\rangle\to|\mp,n{-}1\rangle$. The seven-line collective multiplet is the cooperative generalization of this structure: the additional dressed states introduced by the dipole--dipole coupling $\Omega_{12}$ open new inter-manifold transitions, splitting each Mollow sideband and generating the richer multiplet seen in Fig.~\ref{fig2_CW}(d). In the limit $\Omega_{12}\to 0$, the extra transitions become degenerate with the Mollow lines, and the seven-line structure collapses back toward the triplet; in the single emitter limit $N=1$, the Mollow triplet is recovered.

At strong driving with $\Delta=-\Omega_{12}$ (single-photon-resonance), fewer prominent peaks (Fig.~\ref{fig2_CW}(d), blue curve)  are resolved than in the two-photon case (Fig.~\ref{fig2_CW}(b), blue curve): because the drive is concentrated on the $|G\rangle\leftrightarrow|S\rangle$ branch, leading to a strong redistribution of transition amplitudes. Consequently, several dressed-state transitions contribute only weakly to the fluorescence spectrum.

The CW spectra recover the expected limiting regimes. For weak coupling, $\lvert\Omega_{12}\rvert\ll\Omega_{0}$, and strong driving, the spectrum approaches the single-emitter Mollow triplet, with sidebands near $\omega\simeq\pm\Omega_{0}$~\cite{mollow1969power}. For strong coupling, $\lvert\Omega_{12}\rvert\gg\Omega_{0}$, the dipole-dipole interaction splits the symmetric and antisymmetric collective states, producing cooperative features near $\omega\simeq\pm\Omega_{12}$~\cite{rudolph1995two}. When $\lvert\Omega_{12}\rvert\sim\Omega_{0}$, neither limit applies, and the spectrum develops the characteristic multi-line structure of strongly driven interacting emitters, including the seven-line spectrum~\cite{darsheshdar2021photon,rudolph1995two}. Thus, Eq.~(\ref{E_spectrum}) consistently captures the single-emitter and cooperative resonance-fluorescence limits.

\begin{figure}[h!]
\centering
\includegraphics[height=70mm,width=\linewidth]{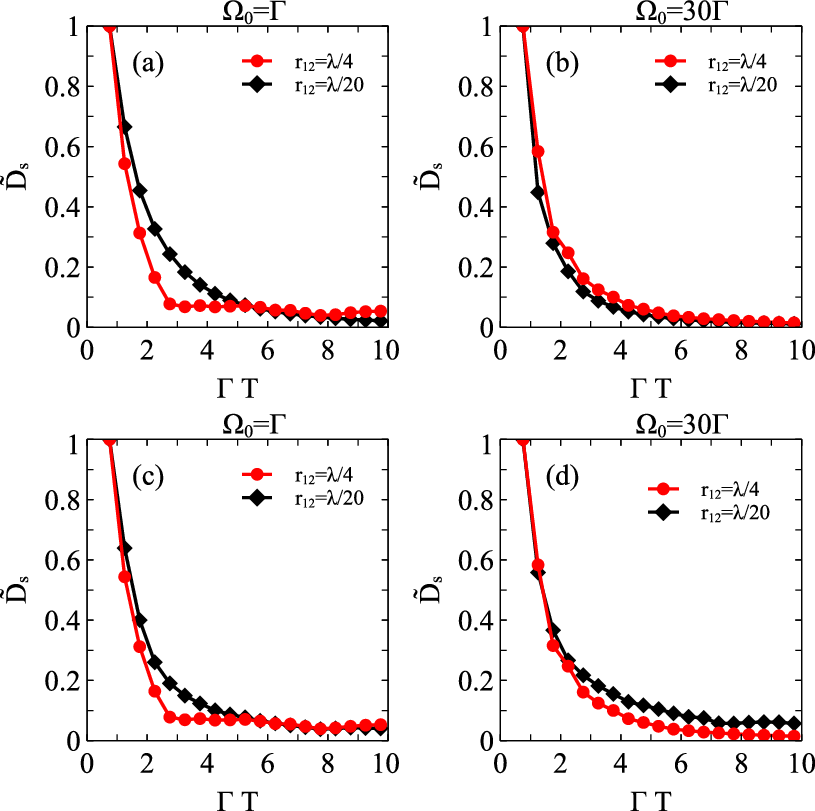}
\caption{Normalized spectral distance $\tilde{D}_{\mathcal{S}}$, tracking the relative change between spectra at successive times, for a continuously driven system of two identical coupled emitters ($\Delta_1 = \Delta_2$) with $\hat{\mu} \perp \hat{r}_{12}$. Panels (a) and (b) correspond to two-photon resonance ($2\omega_L = 2\omega_0$) at Rabi frequencies $\Gamma$ and $30\Gamma$, respectively, while panels (c) and (d) correspond to driving at the superradiant transition frequency ($\Delta=-\Omega_{12}$),  at Rabi frequencies $\Gamma$ and $30\Gamma$, respectively. In each case, results are shown for weak ($r_{12} = \lambda/4$) and strong ($r_{12} = \lambda/20$) dipole-dipole coupling, where $r_{12}$ is the interatomic separation and $\lambda$ is the resonant wavelength.}
\label{SD_CW}
\end{figure}

\subsection{Approach to the Stationary Spectrum}
The spectra discussed so far are transient: computed over a finite observation window $T$, they retain a dependence on $ T$ that the stationary spectrum does not. The unequal sideband heights visible in Fig.~\ref {fig2_CW} are one signature of this non-stationarity. It is therefore natural to ask how, and how quickly, the transient spectrum approaches its stationary form as T increases.

As the total time $T$ increases, the spectral features sharpen. For instance, in the strong coupling and moderate driving regime, the central peak at $\omega = 0$ remains unresolved, and the sidebands at $\pm \Omega_{12}$ are asymmetric in the early transient regime. With increasing $T$, the sideband asymmetry is suppressed, and their amplitudes decrease, whereas the central peak sharpens towards its steady-state Lorentzian profile.

To quantify this convergence across all coupling-driving regimes, we define the normalized spectral distance $\widetilde{D}_\mathcal{S}(T)$
%
%
\begin{equation}
    \widetilde{D}_\mathcal{S}(T) = \frac{\sum_k \left|\bar{\mathcal{S}}(\omega_k; T+\Delta T)
    - \bar{\mathcal{S}}(\omega_k; T)\right|\Delta\omega}{\max_{T}\sum_k
    \left|\bar{\mathcal{S}}(\omega_k; T+\Delta T) - \bar{\mathcal{S}}(\omega_k; T)\right|
    \Delta\omega},
    \label{spectral_distnace_cw}
\end{equation}
where $\bar{\mathcal{S}}(\omega;T) \equiv S(\omega;T)/T$ is the time-normalized spectrum
and $\Delta T$ is the spacing between successive observation times. A large $\tilde{D}_S$ indicates rapid spectral evolution, while a small $\tilde{D}_S$ signals that the spectrum has nearly stabilized. We plot $\tilde{D}_S$ versus $T$ for all coupling-driving regimes to characterize and compare their transient dynamics, which are depicted in Fig~\ref{SD_CW}. The dynamics of $\tilde{D}_s$ in various coupling-driving regimes show that, as the total evolution time increases, the spectral distance between adjacent time steps tends to vanish, and the spectrum becomes progressively more stable.


\section{Collective emission spectra across coupling regimes under pulsed driving}
In this section, we analyze emission spectra generated by periodic pulses and compare them with those produced under continuous driving fields. For a single two-level emitter, a periodic train of instantaneous $\pi$ pulses produces an emission spectrum with a central feature at the carrier frequency and pulse-induced sidebands displaced by multiples of $\pi/\tau_D$~\cite{fotso2018controlling}.
In the following, we show how this structure is modified in a dipole-dipole-coupled two-emitter system under periodic instantaneous $\pi_x$ pulses and identify the conditions under which finite-duration pulses closely reproduce the instantaneous-pulse spectrum.

\subsection{Emission Spectra Under Instantaneous Pulsed Driving}
Here, we consider a periodic train of global $\pi$ pulses applied simultaneously to both emitters about the $x$ axis at times $t_n = n\tau_D$ ($n = 1, 2, \dots$), where $\tau_D$ is the inter-pulse delay. For strong, short pulses satisfying $\Omega_0 \gg \Gamma, |\Delta|, \Omega_{12}$ at fixed pulse area, the pulses can be treated within the instantaneous-pulse approximation, in which each pulse acts as the unitary rotation
\begin{equation}
U_x = \exp\!\left[-i\frac{\pi}{2}\left(\sigma_x\otimes I + I\otimes\sigma_x\right)\right],
\label{instant_pi}
\end{equation}
so that $\rho(t_n^+) = U_x\,\rho(t_n^-)\,U_x^\dagger$. Between pulses, the driving field is off, and the system evolves freely under the master equation~(8) with $\Omega_0 = 0$. The emitters are initialized in $|E\rangle$, and the transient emission spectrum is computed from Eq.~(13) over the time window $0 \le t \le T$ with $\Gamma T = 4$.



As in the single-emitter case, the pulse train also generates sidebands.
Here, however, their positions 
are governed by the interplay between the pulse-induced spectral spacing and the collective level splitting generated by the dipole-dipole interaction.
For emission through the bright collective ladder, $|E\rangle \rightarrow |S\rangle \rightarrow |G\rangle$, the collective interaction shifts the dominant emission channels away from the laser frequency, while the pulse sequence generates spectral replicas. The main spectral features are therefore expected at  
%
\begin{equation}
\omega \simeq \pm \Omega_{12}+n\frac{\pi}{\tau_D},
\qquad n=0,\pm1,\pm2,\ldots ,
\label{eqn_pulse_driving}
\end{equation}
where, $\Omega_{12}$ sets the collective linecenters, and $\pi/\tau_D$ determines the spacing between pulse-induced replicas. This approximate relation serves as a guide for interpreting the evolution of the spectra as a function of inter-pulse delay.

\begin{figure}[t!]
	\includegraphics[height=70mm,width=1\linewidth]  {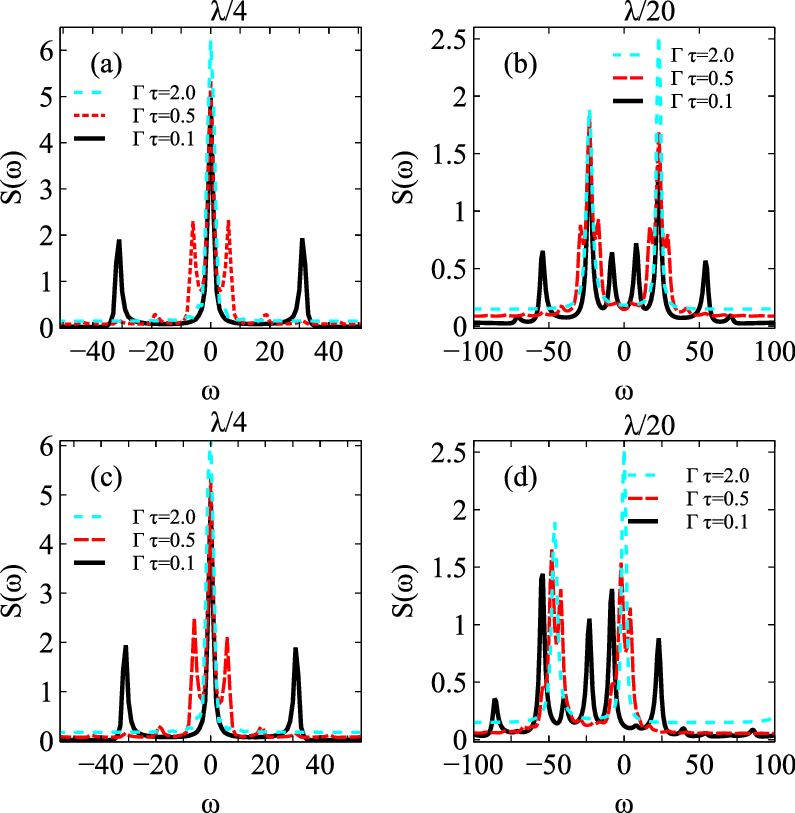}
	\caption{Collective emission spectrum from a coupled two identical ($\Delta_1=\Delta_2$) emitter system under global $\sigma_x$ rotation for $\hat\mu\perp \hat{r}_{12}$. In  (a) and (b), the laser is tuned to two-photon resonance ($2\omega_0=2\omega_L$), and in (c) and (d), the laser is tuned to the transition frequency of the superradiant state ($\Delta=-\Omega_{12}$). The rotation around the $x$ axis is performed at inter pulse intervals $\tau=0.1$, $0.5$ and $2.0$ for dipole-dipole coupling strengths corresponding to $r_{12} = \lambda/4$ and $r_{12} = \lambda/20$, respectively.  $r_{12}$ and $\lambda$ denote the interatomic distance and the resonant wavelength, respectively. The total time of evolution is considered to be $\Gamma t=4$.}
	\label{fig3_CW}
\end{figure}

Figure~\ref{fig3_CW} shows the emission spectra of the two coupled emitters under periodic pulsed driving for inter-pulse delays $\Gamma\tau_D = 0.1$, $0.5$, and $2.0$. The left and right columns correspond to the weak ($r_{12} = \lambda/4$, $\Omega_{12} \approx 0.3\Gamma$) and strong ($r_{12} = \lambda/20$, $\Omega_{12} \approx 23\Gamma$) coupling regimes, respectively. The top row corresponds to the two-photon resonance condition ($\Delta = 0$, i.e., $2\omega_L = 2\omega_0$), and the bottom row to the single-photon resonance condition ($\Delta = -\Omega_{12}$). In all cases, the positions of the resolved peaks agree with the frequencies predicted by Eq.~(\ref{eqn_pulse_driving}).

For $\Gamma \tau_D=0.1$, this gives a central peak near $\omega=0$ together with two prominent sidebands near $\omega\simeq\pm\pi/\tau_D$, producing a three-peak structure similar to the weak-coupling, strong-driving CW spectrum (Fig.~\ref{fig2_CW}(b), red curve). As $\Gamma \tau_D$ increases to $0.5$, the pulse-induced spacing $\pi/\tau_D$ decreases, and the sidebands move toward the central peak, increasing their spectral overlap. For $\Gamma\tau_D=2.0>1/\Gamma$, the sideband spacing becomes comparable to the linewidth scale $\Gamma$, and the individual sideband features are no longer clearly resolved.


\begin{figure*}[!t]
	\centering
	
	\begin{minipage}{0.425\textwidth}
		\centering
		\includegraphics[width=\textwidth]{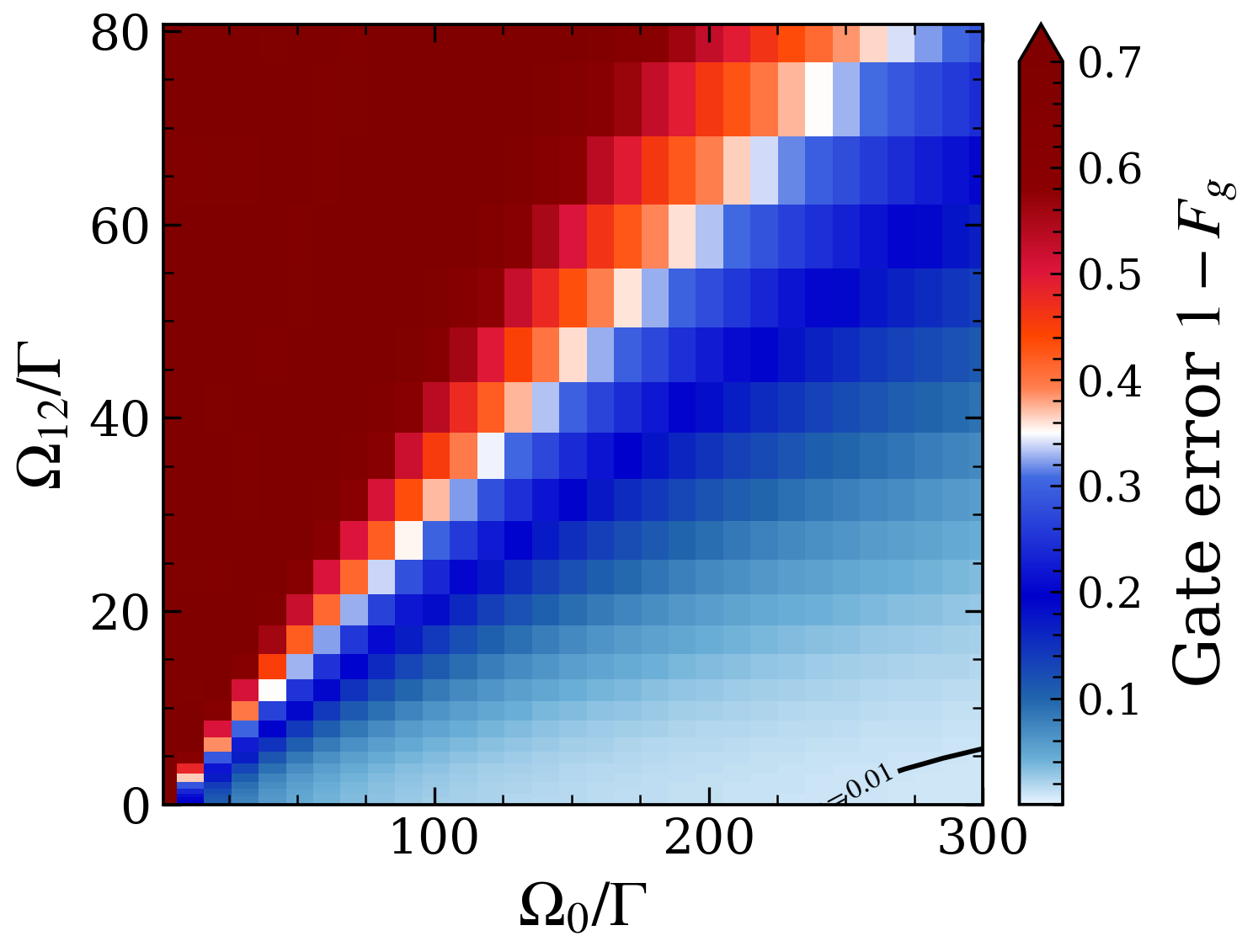}
		\textbf{(a)}
	\end{minipage}
	\hfill
	\begin{minipage}{0.425\textwidth}
		\centering
		\includegraphics[width=\textwidth]{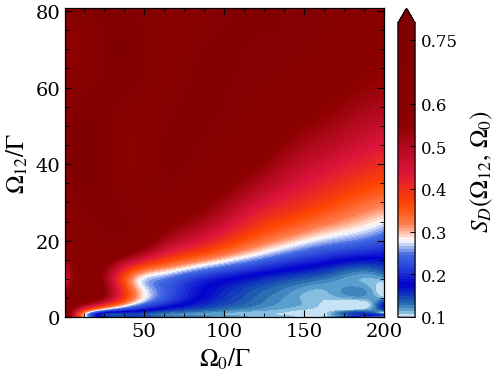}
		\textbf{(b)}
	\end{minipage}
    \hfill

\caption{ (a) Gate error, $1-F_g$, for a finite-duration $\pi$ pulse as a function of the pulse Rabi frequency $\Omega_0/\Gamma$ and the dipole-dipole coupling strength $\Omega_{12}/\Gamma$. The gate error decreases as $\Omega_0$ increases, indicating that sufficiently strong finite-duration pulses closely approximate the ideal instantaneous $\sigma_x$ rotation, even in the presence of inter-emitter coupling. 
(b) Spectral distance, $S_D$, between the spectra generated by finite-duration and instantaneous pulse sequences, evaluated for $N_p=10$ pulses. Smaller values of $S_D$ indicate better agreement with the instantaneous-pulse limit. }
\label{gate_error_SD}
\end{figure*}




In the case of single-photon resonance ($\Delta=-\Omega_{12}$) (Figs.~\ref{fig3_CW}(c) and \ref{fig3_CW}(d)), the spectral features occur at frequencies determined by the dipole-dipole coupling $\Omega_{12}$, with additional pulse-induced sidebands at integer multiples of $\pi/\tau_{D}$. Thus, the peak positions are approximately described by $\omega \simeq \pm\Omega_{12}+n\pi/\tau_D$, while the resonance condition primarily modifies the relative spectral weights of these components. For weak coupling (Fig.~\ref{fig3_CW}(c)), the spectrum remains qualitatively similar to the two-photon-resonance case. 
In the strong-coupling regime [Fig.~\ref{fig3_CW}(d)], by contrast, the redistribution of spectral weight leads to pronounced differences in the relative intensities of the pulse-induced features. The brightest components now lie near $\omega \approx 0$ and $\omega \approx -2\Omega_{12}$, which correspond to the $S \to G$ and $E \to S$ transitions located at $\Delta + \Omega_{12}$ and $\Delta - \Omega_{12}$, respectively, for $\Delta = -\Omega_{12}$. (The $E \to A$ and $A \to G$ transitions are degenerate with these lines, but their contribution is negligible because the antisymmetric state is nearly dark at $r_{12} = \lambda/20$.)

Our results, thus, indicate that under pulsed driving, the line positions are thus determined by $\Omega_{12}$ and $\tau_D$, whereas the resonance condition determines their relative intensities.

\subsection{Finite-Duration Pulses and Gate Error}
In practice, the the $\pi$ pulse in q.~\ref{instant_pi} is implemented by a resonant drive of a finite duration $t_\pi = \pi/\Omega_0$, during which the system evolves under $\Omega_{12}$ and decays. The instantaneous-pulse approximation is therefore valid only when $\Omega_0$ is large compared to other dynamical frequency scales of the system. 
To quantify the performance of a finite-duration $\pi$ pulse as an approximation to the ideal instantaneous $\sigma_x \otimes \sigma_x$ gate, we evaluate the average gate fidelity~\cite{nielsen2002simple}
\begin{equation}
	F_g(\mathcal{E}, U_\pi) = \frac{\sum_j \mathrm{Tr}\!\left( U_\pi U_j^\dagger U_\pi^\dagger \, \mathcal{E}(U_j) \right) + d^2}{d^2(d+1)},
\end{equation}
where $U_\pi = \sigma_x \otimes \sigma_x$ is the target unitary, $d=4$ is the Hilbert space dimension, and $\{U_j\} = \{\sigma_\mu \otimes \sigma_\nu\}$ with $\mu,\nu \in \{0,x,y,z\}$ forms a complete orthonormal operator basis satisfying $\mathrm{Tr}(U_j^\dagger U_k) = d\,\delta_{jk}$. Here, $\mathcal{E}$ denotes the quantum channel generated by a square pulse of duration $t_\pi$, during which the system evolves according to a Lindblad master equation that includes collective decay at rate $\Gamma_{12}$ and dipole-dipole coupling $\Omega_{12}$.

The corresponding gate error is defined as
\begin{equation}
\textrm{	Gate error} = 1 - F_g.
\end{equation}
Fig.~\ref{gate_error_SD}(a) shows the gate error $1 - F_g$ (obtained from finite-time evolution under the full master equation) as function of $\Omega_0$ and $\Omega_{12}$. High-fidelity rotations are attainable only in the strong-driving regime, $\Omega_0 \gg \Omega_{12}$, where the pulse effectively acts instantaneously. When $\Omega_0$ becomes comparable to $\Omega_{12}$, substantial deviations arise, indicating the breakdown of the instantaneous-pulse approximation. Unlike the ideal gate $U_\pi$, which is applied instantaneously, the physical map $\mathcal{E}$ is subject to decoherence throughout the finite pulse duration $t_\pi$. Consequently, the gate error decreases as $\Omega_0$ increases, and hence as $t_\pi$ shortens, vanishing in the limit $\Omega_0\to\infty$.


Because pulse error accumulate over the pulse train, we also compare the emission spectra obtained under an ideal instantaneous $\pi_x$ pulse sequence and under a finite-duration pulse sequence. We quantify their difference by  the spectral distance $\mathcal{S}_D$, 
\begin{equation}
	\mathcal{S}_D(\Omega_{12},\Omega_0)
	=
	\sum_{k=1}^{N_\omega}
	\left|
	\tilde{\mathcal{S}}_{\pi_x}(\omega_k;\Omega_{12})
	-
	\tilde{\mathcal{S}}_{t_\pi}(\omega_k;\Omega_{12},\Omega_0)
	\right|
	\Delta\omega ,
\end{equation}
where
\begin{equation}
	\tilde{\mathcal{S}}(\omega_k)
	=
	\frac{\mathcal{S}(\omega_k)}
	{\sum_{j=1}^{N_\omega} \mathcal{S}(\omega_j)\Delta\omega}
\end{equation}
denotes area-normalized emission spectrum, sampled at $N_\omega$ uniformly spaced frequencies. 
In Fig.~\ref{gate_error_SD}(b), the comparison is carried out for a pulse train with $N_p=10$ pulses and total evolution time $T=N_p\tau$, where $\tau=0.2$ is the inter-pulse delay and $t_\pi=\pi/\Omega_0$ is the finite pulse duration.

By construction, $S_D\in[0,2]$, and $S_D\to0$ when the finite-duration pulse sequence reproduces the ideal instantaneous-$\pi_x$ spectrum. Because the emission spectrum is sensitive to both population dynamics and coherence evolution in the driven-dissipative system, $S_D(\Omega_{12},\Omega_0)$ provides a frequency-domain measure of the
deviation from the instantaneous-pulse spectrum and complements the gate error $\epsilon_g=1-F_g$. Both quantities decrease in the strong-pulse limit $\Omega_0\to\infty$ and increase when $t_\pi$ becomes comparable to $\Gamma^{-1}$, showing that gate infidelity and spectral distortion share the same physical origin: decoherence and unwanted system evolution accumulated during the finite pulse.

\begin{figure}[h!]
	\includegraphics[height=70mm,width=1\linewidth]  {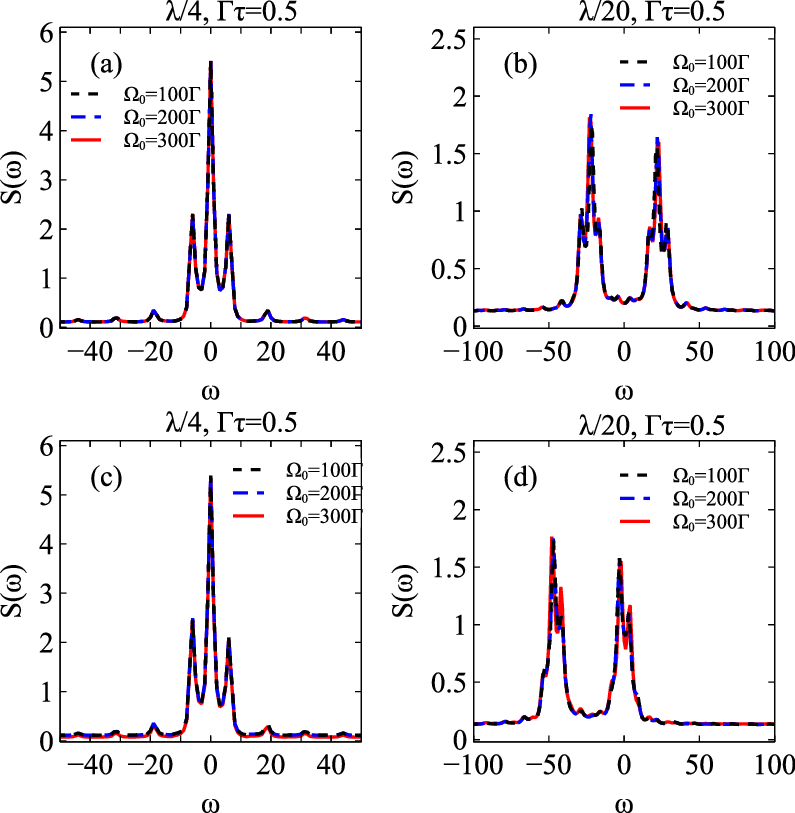}
	\caption{ 
    Collective emission spectra of two identical coupled emitters ($\Delta_1=\Delta_2$, $\hat{\mu}\perp\hat{r}_{12}$) driven by $N_p = 10$ finite-duration $\pi_x$ pulses of Rabi frequency $\Omega_0/\Gamma = 100$, $200$, and $300$ and duration $t_p = \pi/\Omega_0$, applied at intervals $\tau_D\Gamma = 0.5$. In (a) and (b), the laser is tuned to two-photon resonance ($\omega_L = \omega_0$, $\Delta = 0$); in (c) and (d), it is tuned to the superradiant transition ($\omega_L = \omega_0 + \Omega_{12}$, $\Delta = -\Omega_{12}$). Panels (a) and (c) correspond to the weak-coupling regime ($r_{12} = \lambda/4$), and (b) and (d) to the strong-coupling regime ($r_{12} = \lambda/20$), where $r_{12}$ is the interatomic distance and $\lambda$ the resonant wavelength.}
	\label{finite_pulse}
\end{figure}
While $\mathcal{S}_D$ quantifies the overall deviation from the instantaneous-pulse limit, the individual spectra show how this limit is recovered. Figure~\ref{finite_pulse} shows the emission spectra for finite-duration pulses at $\Gamma\tau_D = 0.5$ and $\Omega_0/\Gamma = 100$, $200$, and $300$, for the same four configurations as Fig.~\ref{fig3_CW}: weak (left column) and strong (right column) coupling, at
two-photon (top row) and single-photon (bottom row) resonance. At weak coupling (Figs.~\ref{finite_pulse}(a) and \ref{finite_pulse}(c)), the spectra retain the three-peak structure of the instantaneous-pulse case, a central line at the laser frequency flanked by pulse-induced sidebands, and the two resonance conditions give nearly identical results because the collective splitting is unresolved. At strong coupling (Figs.~\ref{finite_pulse}(b) and \ref{finite_pulse}(d)), the collective splitting separates the emission into two groups of evenly spaced replicas, placed symmetrically about the laser frequency at two-photon resonance and asymmetrically at single-photon resonance. The line positions are fixed by the pulse periodicity, so the finite pulse duration alters only the relative line strengths, most noticeably at strong coupling, where the system evolves appreciably during each pulse. As $\Omega_0$ increases, each pulse approaches a sudden rotation, and the spectra in all four configurations converge to the instantaneous-pulse results of Fig.~\ref{fig3_CW}, consistent with the decrease in gate error and spectral distance. Thus, for a given inter-pulse delay, sufficiently fast pulses recover the spectral features predicted by the instantaneous-pulse approximation.

\section{Photon-photon correlations}

Beyond the emission spectrum, photon–photon correlations provide an equally important perspective on quantum control. Whereas the spectrum characterizes the frequency distribution of the emitted radiation, photon-correlation measurements reveal the temporal statistics, the correlations between successive emission events, and their evolution under continuous and periodic pulse driving. They therefore provide additional insight into the underlying driven-dissipative dynamics, including the emergence of suppressed and enhanced correlations during the transient evolution.

To characterize the temporal quantum statistics of the emitted radiation, we calculate the normalized second-order correlation function
\begin{equation}
	g^{(2)}(t,t+\tau)=
	\frac{\langle C^{\dagger}(t)\,C^{\dagger}(t+\tau)\,C(t+\tau)\,C(t)\rangle}
	{\langle C^{\dagger}(t)\,C(t)\rangle\,\langle C^{\dagger}(t+\tau)\,C(t+\tau)\rangle},
\end{equation}
which we write compactly as
\begin{equation}
	g^{2}(t,\tau)\equiv g^{(2)}(t,t+\tau)=\frac{I^{(2)}(t+\tau,t)}{I(t)\,I(t+\tau)}.
\end{equation}
Here, $(I(t)=\langle C^{\dagger}(t)C(t)\rangle)$ denotes the photon flux, while $(I^{(2)}(t,t+\tau)=\langle C^{\dagger}(t)\,C^{\dagger}(t+\tau)\,C(t+\tau)\,C(t)\rangle)$ represents the joint detection probability at times \(t\) and \(t+\tau\). The collapse operator \(C\) corresponds to the detected emission channel, which is taken as \(C=\sigma_1^-+\sigma_2^-\) in our model. Throughout this section, we focus on the bright radiative pathway \(\ket{E}\!\to\!\ket{S}\!\to\!\ket{G}\), which accounts for the dominant contribution to the emitted intensity.


Following photon detection at time $t$, the system is projected onto the unnormalized conditional state $\widetilde{\rho}(t)=C\rho(t)C^\dagger$ and subsequently evolves under the full Liouvillian $\mathcal{L}$ for a delay $\tau$, expressed as
\begin{equation}
	\widetilde{\rho}(t+\tau)=e^{\mathcal{L}\tau}\,\widetilde{\rho}(t).
\end{equation}
The joint detection probability is then given by
\begin{equation}
	I^{(2)}(t,t+\tau)=
	\operatorname{Tr}\!\left[C^{\dagger}C\,\widetilde{\rho}(t+\tau)\right].
\end{equation}

We present the results as two-dimensional heatmaps of $g^{(2)}(t,\tau)$ in the $(t,\tau)$ plane, which reveal both the transient photon statistics during the excitation pulse and regimes of correlated and anti-correlated emission across delay time $\tau$.
%
%
\begin{figure*}[ht]
	\centering
	
	\begin{minipage}{0.24\textwidth}
		\centering
		\includegraphics[width=\textwidth]{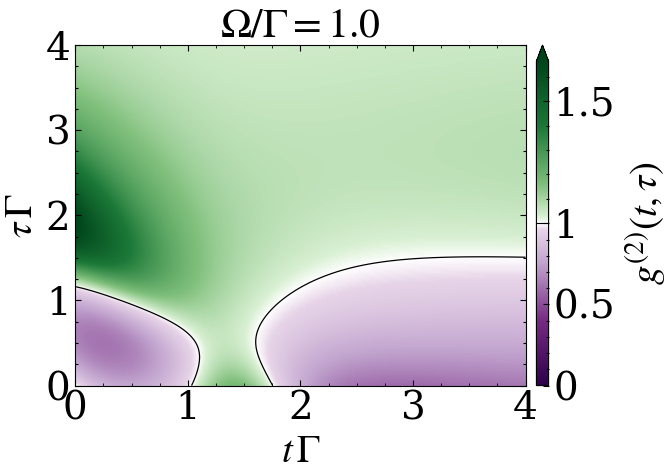}
		\textbf{(a)}
	\end{minipage}
	\hfill
	\begin{minipage}{0.24\textwidth}
		\centering
		\includegraphics[width=\textwidth]{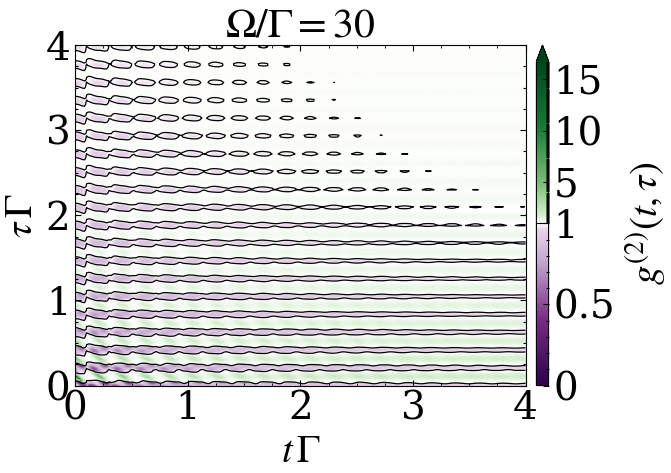}
		\textbf{(b)}
	\end{minipage}
	\hfill
	\begin{minipage}{0.24\textwidth}
		\centering
		\includegraphics[width=\textwidth]{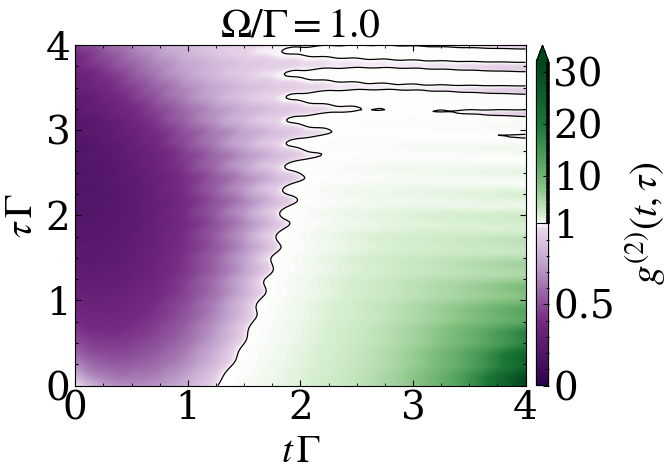}
		\textbf{(c)}
	\end{minipage}
	\hfill
	\begin{minipage}{0.24\textwidth}
		\centering
		\includegraphics[width=\textwidth]{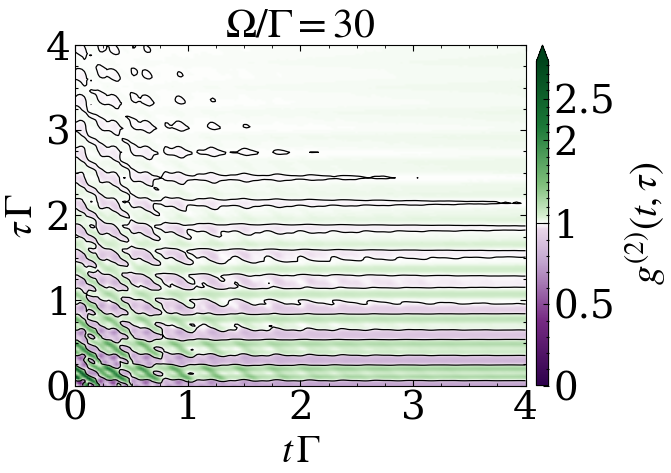}
		\textbf{(d)}
	\end{minipage}
	\caption{Two-time correlation function $g^{(2)}(t,\tau)$ for the cascaded emission pathway $E \rightarrow S \rightarrow G$ in the transient regime under continuous driving. Results are shown at two-photon resonance for dipole moments oriented perpendicular to the interatomic axis, $\hat{\mu} \perp \hat{r}_{12}$. Panels (a) and (b) correspond to the weak-coupling regime($r_{12} = \lambda/4$), under moderate ($\Omega = \Gamma$) and strong ($\Omega = 30\Gamma$) driving, respectively. Panels (c) and (d) show the corresponding results for the strong-coupling regime $(r_{12} = \lambda/20$), under the same moderate and strong driving conditions.}
	
	\label{g2_two_photon}
\end{figure*}

\begin{figure*}[!htb]
	\centering
	
	\begin{minipage}{0.24\textwidth}
		\centering
		\includegraphics[width=\textwidth]{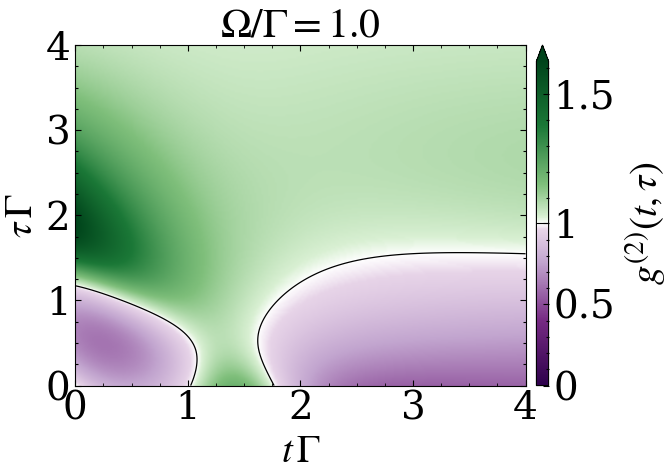}
		\textbf{(a)}
	\end{minipage}
	\hfill
	\begin{minipage}{0.24\textwidth}
		\centering
		\includegraphics[width=\textwidth]{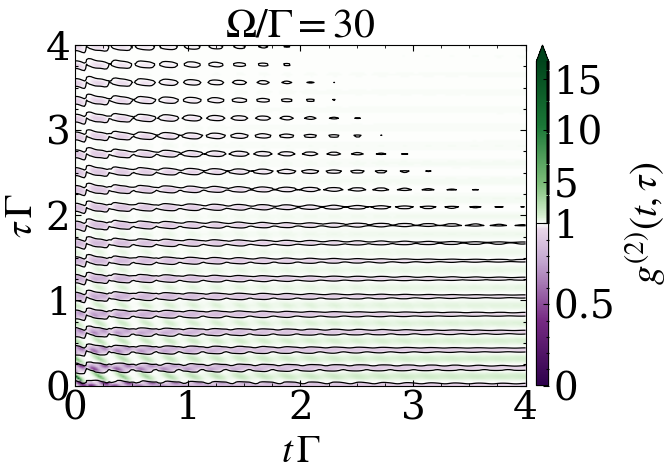}
		\textbf{(b)}
	\end{minipage}
	\hfill
	\begin{minipage}{0.24\textwidth}
		\centering
		\includegraphics[width=\textwidth]{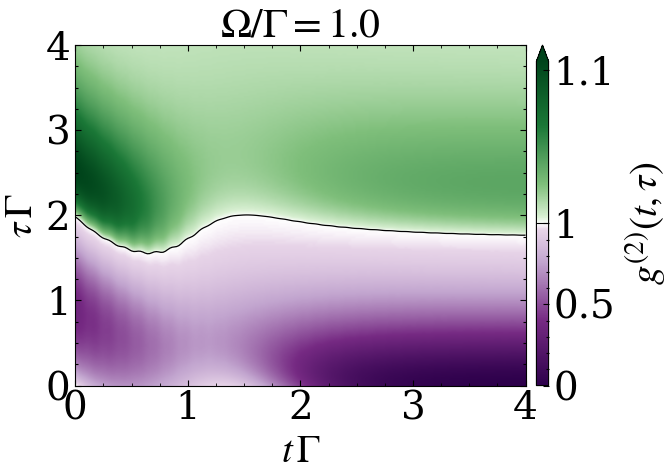}
		\textbf{(c)}
	\end{minipage}
	\hfill
	\begin{minipage}{0.24\textwidth}
		\centering
		\includegraphics[width=\textwidth]{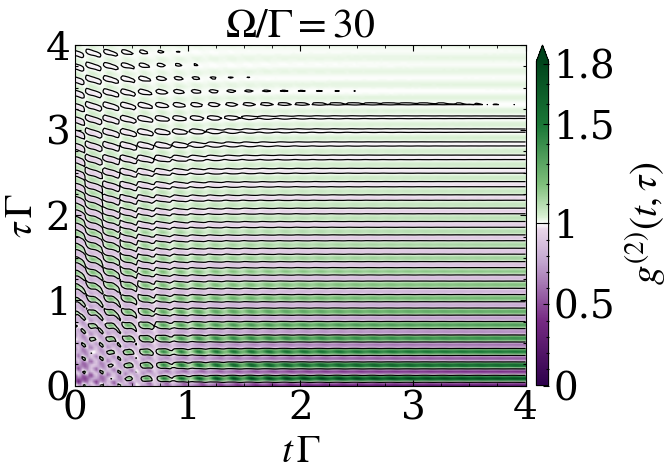}
		\textbf{(d)}
	\end{minipage}
	\caption{Two-time correlation function $g^{(2)}(t,\tau)$ for the cascaded emission pathway $E \rightarrow S \rightarrow G$ in the transient regime under continuous driving. Results are shown at one-photon resonance for dipole moments oriented perpendicular to the interatomic axis, $\hat{\mu} \perp \hat{r}_{12}$. Panels (a) and (b) correspond to the weak-coupling regime($r_{12} = \lambda/4$), under moderate ($\Omega = \Gamma$) and strong ($\Omega = 30\Gamma$) driving, respectively. Panels (c) and (d) show the corresponding results for the strong-coupling regime $(r_{12} = \lambda/20$), under the same moderate and strong driving conditions.
	} 
	\label{g2_one_photon}
	
\end{figure*}

Figures~\ref{g2_two_photon} and \ref{g2_one_photon} illustrate the two-time second-order correlation function $g^{(2)}(t,\tau)$ under continuous-wave (CW) driving across different driving and coupling regimes at the two-photon and single-photon resonances, respectively. In the weak-coupling, moderate-driving regime, the qualitative behavior of $g^{(2)}(t, \tau)$ is strikingly similar at both resonances (Figs.~\ref{g2_two_photon}(a) and \ref{g2_one_photon}(a)): the emission is anticorrelated  ($g^{(2)} < 1$) at small $t$ and $\tau$, crossing into correlated emission ($g^{(2)} > 1$) as both $t$ and $\tau$ increase, with the Poissonian boundary $g^{(2)} = 1$ clearly delineated by the black contour. In the weak-coupling, strong-driving regime (Figs.~\ref{g2_two_photon}(b) and \ref{g2_one_photon}(b)), rapid Rabi oscillations dominate the dynamics, producing a pronounced oscillatory pattern in both $t$ and $\tau$, with the emission alternating between correlated and anticorrelated regimes. In this regime, the two-time correlation functions at both single-photon and two-photon resonances exhibit a similar qualitative trend. The correlation function reaches significantly higher peak values compared to the moderate-driving case.

In Figs.~\ref{g2_two_photon}(c) and \ref{g2_one_photon}(c), we show $g^{(2)}(t, \tau)$ in the strong-coupling, moderate-driving regime at two-photon and single-photon resonances, respectively. The two cases exhibit markedly different behavior. At two-photon resonance, $g^{(2)}(t, \tau)$ evolves from an anticorrelated to correlated regime as $t$ increases, with the $g^{(2)} = 1$ boundary forming an oscillatory, nearly vertical contour. This reflects coherent Rabi-like cycling between $|gg\rangle$ and $|ee\rangle$ via the two-photon pathway, with the correlation reaching values as large as $g^{(2)} \sim 32$ due to cascaded pair emission from the doubly-excited state. At single-photon resonance, by contrast, the transition is governed by the delay time $\tau$ rather than $t$: $g^{(2)} < 1$ for small $\tau$ and $g^{(2)} > 1$ for large $\tau$, with a smooth horizontal contour near $\tau \Gamma \approx 1.8$ 
and deeply anti-correlated values $g^{(2)} \sim 0.01$ 
at short delays, consistent with strongly quantum single-photon emission from the superradiant state with no pair-emission enhancement. In the strong-coupling, strong-driving regime (Figs.~\ref{g2_two_photon}(d) and \ref{g2_one_photon}(d)), 
the interplay between collective dipole-dipole coupling and intense driving produces qualitatively distinct dynamics at the two resonances. At two-photon resonance, $g^{(2)}(t,\tau)$ exhibits strongly non-stationary behavior, with diagonal oscillatory fringes reflecting transient cascade dynamics-the system undergoes rapid coherent cycling through the doubly-excited state, and the photon statistics retain a strong memory of the initial time $t$ throughout the evolution. At single-photon resonance, the behavior is qualitatively different: $g^{(2)}(t,\tau)$ becomes nearly stationary, displaying regular horizontal oscillations along $\tau$ with little dependence on $t$. This contrast highlights how two-photon resonance sustains genuinely 
two-time, non-stationary correlations even under strong driving. In contrast, single-photon resonance rapidly settles into a quasi-stationary oscillatory regime controlled primarily by the delay time $\tau$.

\begin{figure}[!htb]
    \centering

    \begin{minipage}{0.48\columnwidth}
        \centering
        \includegraphics[width=\linewidth]{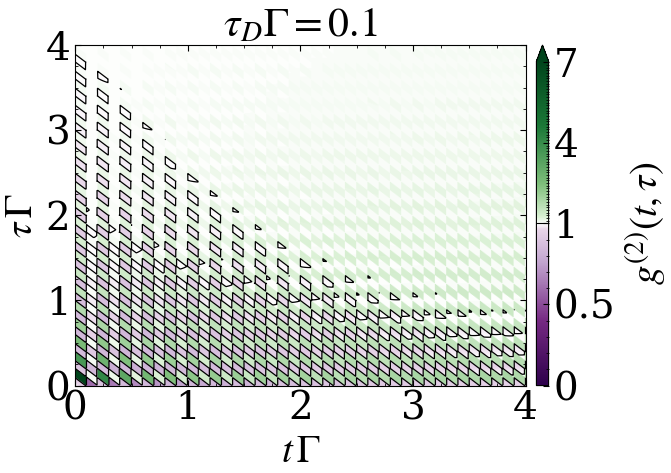}
        \textbf{(a)}
    \end{minipage}
    \hfill
    \begin{minipage}{0.48\columnwidth}
        \centering
        \includegraphics[width=\linewidth]{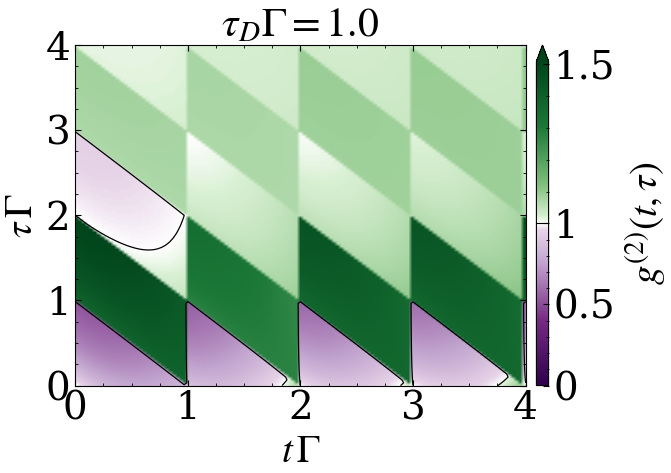}
        \textbf{(b)}
    \end{minipage}

    \vspace{0.5em}

    \begin{minipage}{0.48\columnwidth}
        \centering
        \includegraphics[width=\linewidth]{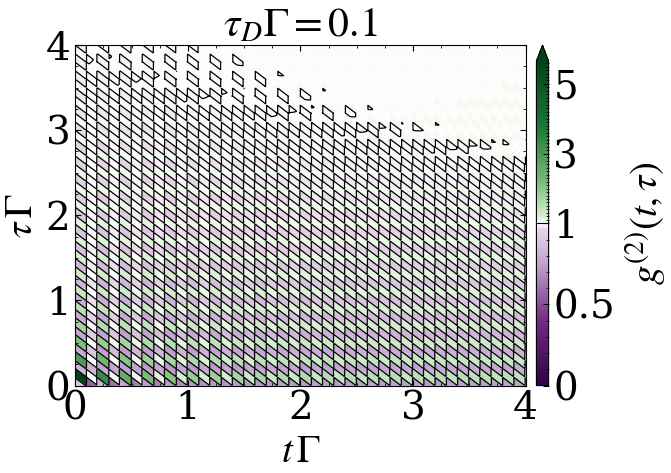}
        \textbf{(c)}
    \end{minipage}
    \hfill
    \begin{minipage}{0.48\columnwidth}
        \centering
        \includegraphics[width=\linewidth]{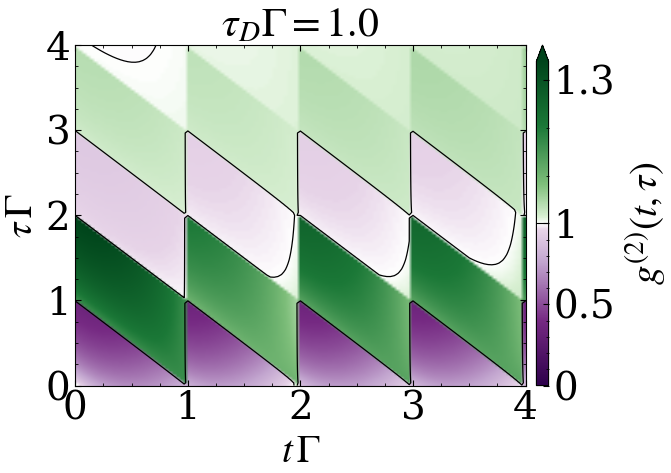}
        \textbf{(d)}
    \end{minipage}

    \caption{The two-time correlation function, $g^{(2)}(t,\tau)$, is shown for cascaded emission ($E \rightarrow S \rightarrow G$) in the transient regime under pulsed driving (global $\sigma_x$ rotation) at two-photon resonance, with $\hat{\mu} \perp \hat{r}_{12}$. Panels (a) and (b) correspond to the weak-coupling regime ($r_{12}=\lambda/4$) with inter-pulse delays $\tau_D = 0.1$ and $\tau_D = 1.0$, respectively, while panels (c) and (d) show the strong-coupling regime ($r_{12}=\lambda/20$) for $\tau_D = 0.1$ and $\tau_D = 1.0$, respectively.}
    \label{two_photon_sigma_x}
\end{figure}

In Fig.~\ref{two_photon_sigma_x}, we present the two-time correlation function $g^{(2)}(t,\tau)$ in the weak-coupling (Figs.~\ref{two_photon_sigma_x}(a), and (b)) and strong-coupling (Figs.~\ref{two_photon_sigma_x}(c), and (d)) regimes under pulsed driving (shown for two-photon resonance, $\Delta=0$), for two inter-pulse delays, $\tau_D\Gamma = 0.1$ and $1.0$. The emitters, initially in $|E\rangle$, are driven by instantaneous global $\pi_x$ pulses (Eq.~(\ref{instant_pi})) applied at $t = n\tau_D$, and photons are detected in the bright channel $C = \sigma_1^- + \sigma_2^-$. Because pulses are instantaneous, $g^{(2)}(t,\tau)$ is piecewise smooth, with discontinuities along $t = n\tau_D$ and $t+\tau = m\tau_D$; each region is labeled by the number of pulses preceding the first detection and separating the two detections.

To interpret these features, we first note that the correlations are insensitive to the coherent coupling and the detuning. Between pulses the emitters evolve freely ($\Omega_R = 0$), so that in the collective basis $\{|G\rangle, |S\rangle, |A\rangle, |E\rangle\}$ the Hamiltonian is diagonal, $H = \mathrm{diag}(-\Delta, \Omega_{12}, -\Omega_{12}, \Delta)$. The pulse exchanges $|G\rangle \leftrightarrow |E\rangle$ and leaves the single-excitation populations unchanged, decay proceeds through incoherent cascades, and a detection in the symmetric channel maps $E \to S$ and $S \to G$. Starting from $|E\rangle$, both the unconditioned and the conditioned density matrices therefore remain diagonal, $[H,\rho] = 0$, and $g^{(2)}$ does not depend explicitly on $\Omega_{12}$ or $\Delta$; the coupling enters only through the collective decay rate $\Gamma_{12}$. One-photon ($\Delta = -\Omega_{12}$) and two-photon ($\Delta = 0$) resonant driving thus yield identical results, as expected for instantaneous pulses, which lack spectral selectivity.

The two coupling regimes therefore differ only through the collective decay, that is, through the superradiant rate $\Gamma+\Gamma_{12}$ of the bright cascade $E\to S\to G$. A first detection projects the pair onto $|S\rangle$, which is unaffected by the subsequent pulse, while the population that has decayed to $|G\rangle$ is transferred back to
$|E\rangle$; the entire population thus remains in the bright cascade, and the conditioned emission after the pulse is maximal and coupling independent. In the unconditioned (average) dynamics, the pulse transfers the population still in $|E\rangle$ to the non-emitting state $|G\rangle$, so that only the population that has left $|E\rangle$ during one period continues to emit after the pulse.
For short inter-pulse delays, this population is small, the unconditioned emission after the pulse is weak, and the ratio of conditioned to unconditioned emission becomes large, reaching $g^{(2)} \approx 7$ and $\approx 5.6$ for $\tau_D\Gamma = 0.1$ in the weak- and strong-coupling regimes ( Figs.~\ref{two_photon_sigma_x} (a) and (c)), respectively. The larger $\Gamma_{12}$ in the strong-coupling regime enhances superradiance and increases the unconditioned emission, which reduces $g^{(2)}$. For $\tau_D\Gamma = 1$, most of the population completes the cascade within a single period, and the correlations remain weak ($g^{(2)} \lesssim 1.5$) in both regimes ( Figs.~\ref{two_photon_sigma_x} (b) and (d)).

For finite-duration $\pi$ pulses, with peak Rabi frequency $\Omega_0$ and duration $t_p$ satisfying $\Omega_0 t_p = \pi$, this equivalence is expected to hold only in the strong, short-pulse limit. In this limit, $\Omega_0 \gg |\Omega_{12}|, |\Delta|, \Gamma$ (equivalently $t_p \ll 1/|\Omega_{12}|, 1/\Gamma$) and $t_p \ll \tau_D$, so the pulse approximates the instantaneous $\pi_x$ rotation; this condition is most demanding in the strong-coupling regime. For longer or weaker pulses, evolution under $\Omega_{12}$ and $\Delta$ during the pulse would make the excitation pathways, and hence $g^{(2)}(t,\tau)$, sensitive to the resonance condition.

\section{Discussion and Conclusion}
We have investigated the collective fluorescence spectrum and transient photon correlations $g^{(2)}(t,\tau)$ of two coupled emitters under continuous-wave and pulsed excitation. Under CW driving, weak coupling produces spectra resembling those of independent emitters, whereas strong coupling reveals collective-state splitting and resonance-dependent features. Strong driving generates a Mollow-like
structure for weak coupling and a richer spectrum for strong coupling, which is best understood in terms of transitions between atom-field dressed states.

Under periodic $\pi_x$ pulses, the spectrum is governed by the interplay between the dipole-dipole coupling $\Omega_{12}$ and the inter-pulse delay $\tau_D$: the main spectral features are organized at
$\omega\approx\pm\Omega_{12}+n\pi/\tau_D$, while the resonance condition primarily redistributes the spectral weight among these components. For weak coupling, $\Omega_{12}$ falls below the linewidth, the collective splitting is unresolved, and the spectrum reduces to a three-peak structure whose sidebands shift toward the central peak as $\tau_D$ increases. 
For strong coupling, distinct replicas appear
around both collective transition frequencies $\pm\Omega_{12}$; when $\tau_D\sim\pi/\Omega_{12}$ the $n=\mp1$ sidebands overlap near $\omega\approx0$, producing an enhanced central feature, and with increasing $\tau_D$ the sidebands move toward $\pm\Omega_{12}$ and become progressively less resolved. Since $\Omega_{12}$ is fixed by the emitter geometry, the inter-pulse delay provides a convenient, experimentally accessible means of controlling the positions of the pulse-induced sidebands. Gate-error and spectral-distance analyses show that finite-duration pulses accurately reproduce ideal instantaneous rotations when the Rabi frequency is sufficiently large, supporting the experimental feasibility of the scheme.

The photon statistics provide complementary information. Weak coupling produces little dependence on the resonance condition, whereas strong coupling under CW driving generates pronounced 
anticorrelated and correlated emission, particularly at moderate driving and two-photon resonance. Under pulsed excitation, in the instantaneous pulse-limit, $g^{(2)}(t,\tau)$ is governed mainly by the inter-pulse delay and decreases as $\tau_D$ increases. It is strictly independent of the resonance condition, because between pulses $(\Omega_R=0)$ the Hamiltonian is diagonal in the collective basis, so the  detuning does not affect $g^{(2)}(t,\tau)$. The resonance condition thus reshapes the emission spectrum under pulsed driving while leaving the photon statistics unchanged.

These results should be experimentally accessible in nanophotonic quantum-dot systems and superconducting circuits, which support strong emitter–emitter interactions, resonance-fluorescence measurements, and fast coherent control. For example, quantum dots coupled to nanophotonic waveguides have demonstrated photon-mediated collective emitter-emitter interactions, leading to superradiant emission and super-/subradiant ynamics~\cite{kim2018super,tiranov2023collective}. The strong-driving regime, $\Omega_0\gg\Gamma$, is likewise well established in resonance-fluorescence experiments on single quantum dots through the observation of Mollow-triplet emission~\cite{flagg2009resonantly,nick2009spin}. Moreover, ultrafast optical pulses have been used to coherently control quantum-dot spins and excitons on picosecond timescales~\cite{press2008complete,wigger2018rabi}. Superconducting circuits provide a complementary platform for implementing strong, tunable inter-qubit interactions and fast microwave control pulses with high precision~\cite{li2020tunable}. In these systems, microwave scattering and fluorescence experiments provide direct access to the radiative decay and decoherence rates of superconducting artificial atoms, and can resolve resonance-fluorescence features such as Mollow-type  spectra~\cite{lu2021characterizing}. The same scheme can also be extended to more than two emitters, to nonidentical emitters for which the antisymmetric state is no longer dark, or to sequences combining rotations about
several axes, which would act on the collective coherence rather than
only on the populations.

Our results therefore show that the two driving schemes offer complementary control over collective emission. Under continuous-wave (CW) excitation, the spectrum and the photon statistics are governed jointly by the drive strength and the detuning: the dressed-state structure of the driven pair fixes both the multiplet of emission lines and the correlations between successive photons. Under periodic pulsed driving, these two aspects separate. The inter-pulse delay $\tau_D$ positions the pulse-induced sidebands around the collective transitions and governs the photon statistics, whereas the resonance condition
redistributes spectral weight but, in the instantaneous-pulse limit, leaves $g^{(2)}(t,\tau)$ unchanged. The emission spectrum can therefore be reshaped through the detuning without altering the photon statistics, which are controlled mainly by the pulse timing. 

\section*{Acknowledgement}
KGP and HT are supported by the National Science Foundation under Grant No. QIS-2328752.

\bibliography{main.bib}
\end{document}